\documentclass[twocolumn]{emulateapj}
\usepackage{ amsmath, color, natbib,todonotes,graphicx,color}

\newcommand\ii{{\sc ii}}
\newcommand\iii{{\sc iii}}

\begin{document}
\title{The metallicity of void dwarf galaxies}

\author{K. Kreckel\altaffilmark{1}, K. Croxall\altaffilmark{2}, B. Groves\altaffilmark{1}, R. van de Weygaert\altaffilmark{3}, R. W. Pogge\altaffilmark{2,4}}

\altaffiltext{1}{Max Planck Institut f\"{u}r Astronomie, K\"{o}nigstuhl 17, 69117 Heidelberg, Germany;  kreckel@mpia.de}
\altaffiltext{2}{Department of Astronomy, The Ohio State University, 140 West 18th Avenue, Columbus, OH 43210, USA}
\altaffiltext{3}{Kapteyn Astronomical Institute, University of Groningen, PO Box 800, 9700 AV Groningen, the Netherlands}
\altaffiltext{4}{Center for Cosmology and Astro-Particle Physics, The Ohio State University, Columbus, OH 43201, USA}

\begin{abstract}
The current $\Lambda$CDM cosmological model predicts that galaxy evolution proceeds more slowly in lower density environments, suggesting that voids are a prime location to search for relatively pristine galaxies that are representative of the building blocks of early massive galaxies. To test the assumption that void galaxies are more pristine, we compare the evolutionary properties of a sample of dwarf galaxies selected specifically to lie in voids with a sample of similar isolated dwarf galaxies in average density environments. We measure gas-phase oxygen abundances and gas fractions for eight dwarf galaxies (M$_{r} > -16.2$), carefully selected to  reside within the lowest density environments of seven voids, and apply the same calibrations to existing samples of isolated dwarf galaxies.  
We find no significant difference between these void dwarf galaxies and the isolated dwarf galaxies, suggesting that dwarf galaxy chemical evolution proceeds independent of the large-scale environment.   
While this sample is too small to draw strong conclusions, it suggests that external gas accretion is playing a limited role in the chemical evolution of these systems, and that this evolution is instead dominated mainly by the internal secular processes that are linking the simultaneous growth and enrichment of these galaxies.
\end{abstract}

\section{Introduction}

Void galaxies, found occupying large ($\sim$20 Mpc) underdense void regions, are an environmentally defined population whose isolated nature and extreme environment provides an ideal opportunity to test theories of galaxy formation and evolution (see e.g. \citealt{vdWeygaert2011b} for a review).  They are typically small, gas-rich disk galaxies \citep{Kreckel2012} that are systematically bluer with higher rates of star formation \citep{Grogin1999, Grogin2000}, suggesting they are less evolved than galaxies of comparable mass in higher density environments. This arises naturally from $\Lambda$CDM cosmology, as the reduced matter density in voids results in a longer time between mergers  and thus slower evolution \citep{Goldberg2004}.   To date, some of the lowest metallicity galaxies have been discovered within void environments \citep{Pustilnik2010, Pustilnik2013}, but the connection with the large-scale environment is not clear \citep{Nicholls2014a}.

Cosmological models of galaxy formation predict that void galaxies should be younger and thus more metal poor compared to `field' populations of dwarf galaxies \citep{Goldberg2004}. Previous work studying galaxies in the nearby Lynx-Cancer void show systematically lower metallicities for a range of luminosities \citep{Pustilnik2011}, with significant ($\sim$0.15 dex) offset to lower metallicities for the dwarf galaxies in their sample.  However, the Lynx-Cancer void galaxy study is restricted to a single nearby void environment, and at a distance of 18 Mpc it is more difficult to constrain the extent of the void and reliably determine true void membership.    
Here we measure the gas-phase oxygen abundances of eight H~\textsc{i} rich dwarf galaxies in seven voids using MODS1 at the LBT.  
This sample allows us to test our understanding of the effects of the void environment on the evolution of dwarf galaxies, and provides constraints on the gas content and metallicity for this population that must be matched by simulations.

\section{Sample and Data}

\begin{figure*}[t!]
\includegraphics[height=1.45in]{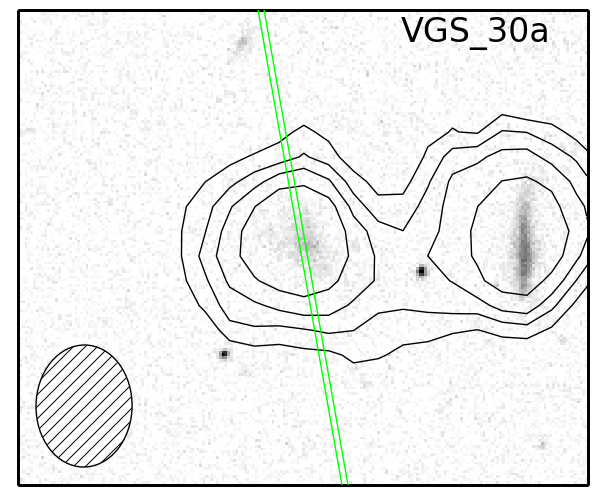}
\includegraphics[height=1.45in]{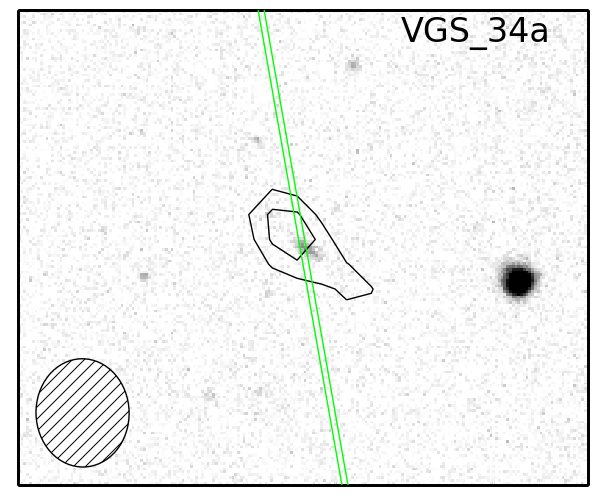}
\includegraphics[height=1.45in]{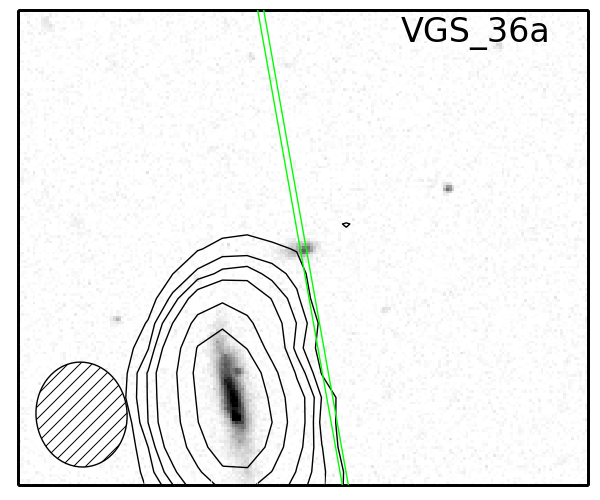}
\includegraphics[height=1.45in]{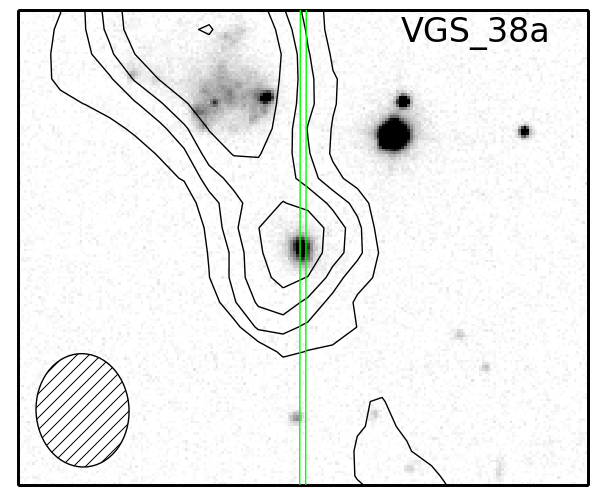}\\
\includegraphics[height=1.45in]{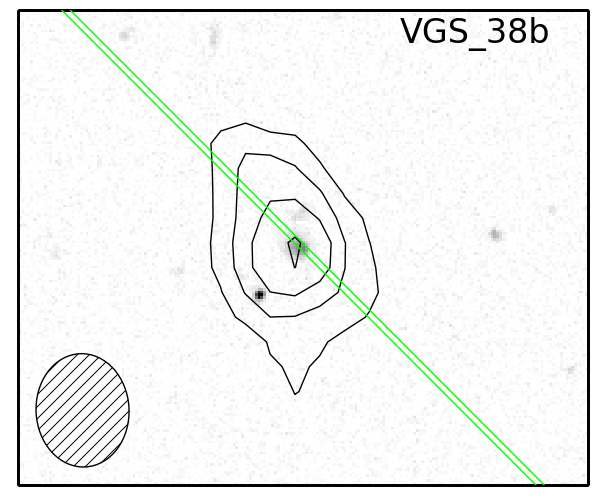}
\includegraphics[height=1.45in]{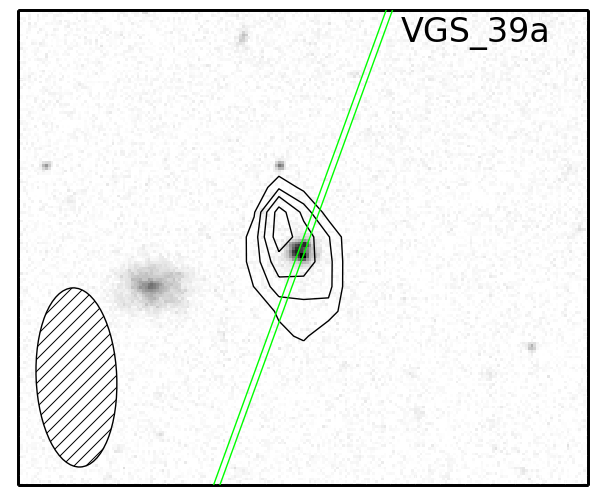}
\includegraphics[height=1.45in]{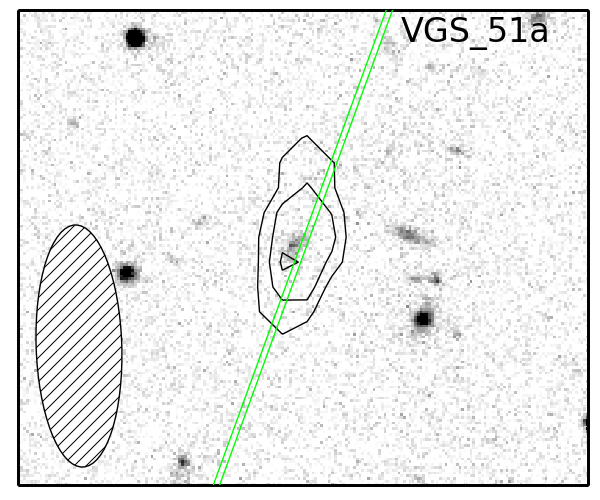}
\includegraphics[height=1.45in]{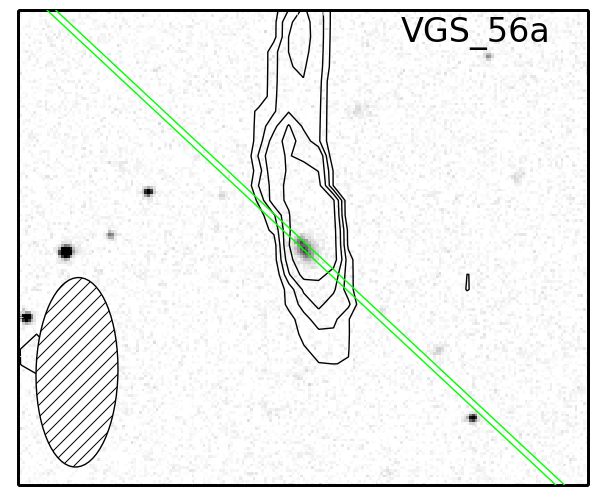}\\
\includegraphics[width=7.3in]{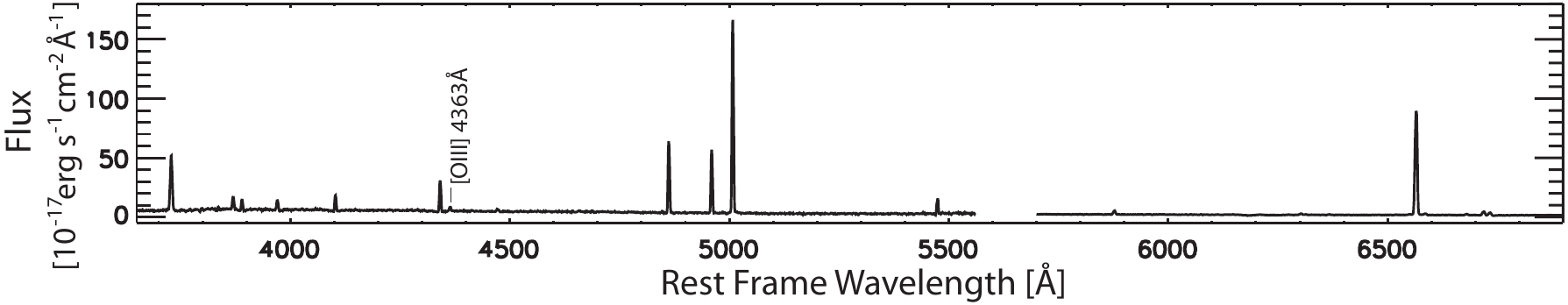}
\caption{Top: SDSS g-band images for the VGS dwarfs, with 48\arcsec field of view.  Contours show the mapped H~\textsc{i} emission from \cite{Kreckel2012} and beam size, and the MODS1 slit position.  VGS\_30a, VGS\_36a and VGS\_38a show signs of interaction with their nearby neighbor in the H~\textsc{i} morphology.  VGS\_38b is also tenuously connected by an extended H~\textsc{i} bridge with VGS\_38a.  
Bottom: The MODS1 spectrum for VGS\_38a shows our detection of the underlying continuum.  Strong emission lines are clearly detected, as well as the faint $\lambda$4363\AA~ line.  
\label{fig:ha}}
\end{figure*}

We observed HII regions in eight galaxies using the MODS1 instrument at the LBT in Spring 2014.  These galaxies were discovered serendipitously within the survey region of an H~\textsc{i} study of void galaxies as part of the Void Galaxy Survey (VGS, \citealt{vdWeygaert2011, Kreckel2012, BeyguPhD}).  The eight dwarf void galaxies reside in seven different voids, with the void boundaries entirely contained within the SDSS redshift survey.
H$\alpha$ imaging, when available, generally reveals only a single HII region \citep{BeyguPhD}.  

\begin{deluxetable*}{l c c c c c c c | c c c c | c}[]
\tablecaption{Galaxy parameters.
\label{tab:properties}}
\tabletypesize{\small}
\tablehead{
\colhead{Name} &
\colhead{ra} &
\colhead{dec} &
\colhead{z} &
\colhead{isolated?} &
\colhead{M$_r$} &
\colhead{M$_B$} &
\colhead{Log[M$_{HI}$]} &
\multicolumn{4}{c}{12+log(O/H)} &
\colhead{log(N/O)}  \\
\colhead{} &
\colhead{(J2000)} &
\colhead{(J2000)} &
\colhead{} &
\colhead{} &
\colhead{(mag)} &
\colhead{(mag)} &
\colhead{(M$_\odot$)} &
\colhead{(PT05)} &
\colhead{(M91)} &
\colhead{(KK04)} &
\colhead{(T$_e$)} &
\colhead{(dex)}
}
\startdata
\multicolumn{3}{l}{VGS dwarf galaxies} \\
\hline
VGS\_30a & 13:05:31 & +54:45:53 & 0.019435 & no & -16.2 & -15.6  & 8.7 & 7.57 & 8.00 & 8.19 & -- & -- \\ 
VGS\_34a & 13:26:41 & +59:32:03 & 0.016539 & yes& -14.0  & -13.5  & 7.7 & 7.74 & 7.92 & 8.14 & -- & --  \\ 
VGS\_36a & 13:55:33 & +59:31:11 & 0.022398 & no & -15.8  &  -14.8        & --    & 7.81 & 7.94 & 8.17 &-- & -1.56 \\ 
VGS\_38a & 14:00:32 & +55:14:46 & 0.013820 & no & -16.1  & -15.9  & 7.9 & 8.05 & 8.18 & 8.38 & 8.01 & -1.45 \\ 
VGS\_38b & 14:00:25 & +55:13:18 & 0.013820 & no & -14.9  & -14.5  & 8.1 & 7.77 & 7.93 & 8.16 & 7.94 & -1.48 \\ 
VGS\_39a & 14:03:21 & +32:45:30 & 0.019037 & yes& -15.2 & -14.9  & 8.4 & 7.45 & 7.60 & 7.89 & 7.54 & -1.46 \\ 
VGS\_51a & 15:12:30 & +24:33:52 & 0.025307 & yes& -14.6  &  -12.0         & 8.0 & 7.60 & 7.75 & 8.01 & -- & --  \\ 
VGS\_56a & 15:34:43 & +28:13:10 & 0.018715 & yes& -15.7  & -14.9  & 8.3 & 7.65 & 8.15 & 8.31 &-- & --  \\   
\hline
\multicolumn{3}{l}{\cite{vanZee2006}} \\
\hline
UGC 12894	& 00:00:22	& +39:29:47	& -- & yes & -- &  -13.4	& --  & 7.60	 & 7.74		& 8.00 & 7.56  & -1.51 \\
UGC 290		& 00:29:08	& +15:54:03	& -- & yes & -- & -14.5	&  -- & 7.60	 & 7.81		& 8.04 & 7.80 & -1.42 \\
UGC 685		& 01:07:22	& +16:41:04	& -- & yes & -- & -14.4	&  -- & 8.01	 & 8.12		& 8.32 & 8.00 & -1.45 \\
UGC 1104    	& 01:32:42	& +18:19:01	& -- & yes & -- & -16.1	&  -- & 8.02	 & 8.11		& 8.30 & 7.94 & -1.65 \\
UGC 1175	& 01:39:57	& +11:05:47	& -- & yes & -- & -14.1	&  -- & 7.51	 & 7.84		& 8.06 & 7.82 & -1.50 \\
UGC 1281	& 01:49:31	& +32:35:17	& -- & yes & -- & -14.9	&  -- & 7.67	 & 7.81		& 8.05 & 7.78 & -1.29 \\
HKK97 L14		& 02:00:10	& +28:49:47	& -- & yes & -- & -11.3	&  -- & 7.52	 & 7.66		& 7.94 & 7.65 & -1.26 \\
UGC 2023    	& 02:33:19	& +33:29:29	& -- & yes & -- & -16.5   	&  -- & 7.86	 & 8.08		& 8.27 & 8.02 & -1.35\\
UGC 3647    	& 07:04:50	& +56:31:10	& -- & yes & -- & -17.1   	&  -- & 7.94	 & 8.06		& 8.28 & 8.07 & -1.28 \\
UGC 3672    	& 07:06:27	& +30:19:21	& -- & yes & -- & -15.4   	&  -- & 7.40	 & 7.79		& 8.02 & 8.01 & -1.64 \\
UGC 4117  	& 07:57:26	& +35:56:25	& -- & yes & -- & -14.9	&  -- & 7.79	 & 7.93		& 8.16 & 7.89 & -1.52 \\
UGC 4483   	& 08:37:03	& +69:46:36	& -- & yes & -- & -12.6	&  -- & 7.51	 & 7.68		& 7.95 & 7.56 & -1.57 \\
CGCG 007-025	& 09:44:02	& -00:38:33	& -- & yes & -- & -15.8	&  -- & 7.72	 & 7.91	& 8.15 & 7.83 & -1.48 \\
UGC 5288   	& 09:51:17	& +07:49:40	& -- & yes & -- & -14.4	&  -- & 8.06	 & 8.13		& 8.32 & 7.90 & -1.42 \\
UGCA 292   	& 12:38:40	& +32:45:49	& -- & yes & -- & -11.4	&  -- & 7.28	 & 7.34		& 7.67 & 7.32 & -1.44 \\
UGC 8651   	& 13:39:53	& +40:44:20	& -- & yes & -- & -13.0	&  -- & 7.66	 & 7.80		& 8.04 & 7.85 & -1.60 \\
UGC 9240   	& 14:24:44	& +44:31:34	& -- & yes & -- & -14.0	&   -- & 7.83	 & 7.98		& 8.21 & 7.95 & -1.60 \\
UGC 9992   	& 15:41:48	& +67:15:14	& -- & yes & -- & -15.0	&  -- & 7.55	 & 7.92		& 8.12 & 7.88 & -1.26 \\
UGC 10445    	& 16:33:48	& +28:59:05	& -- & yes & -- & -17.5   	&  -- & 7.90	 & 8.07		& 8.27 & 7.95 & -1.20 \\
UGC 11755    	& 21:23:00	& +02:24:51	& -- & yes & -- & -17.1   	&  -- & 7.87	 & 7.99		& 8.20 & 8.04 & -1.10 \\
UGC 12713	& 23:38:14	& +30:42:31	& -- & yes & -- & -14.8	&  -- & 7.67	 & 7.83		& 8.07 & 7.80 & -1.53 \\
\hline
\multicolumn{3}{l}{ \cite{Nicholls2014b}} \\
\hline 
J0005-28	& 00:05:32	& -28:05:53	& -- & yes & -- &  -15.0	& 8.23 & 8.00	& 8.10	& 8.31 & 7.85  & -1.68 \\
J1118-17	& 11:18:03	& -17:38:31	& -- & yes & -- &  -13.6	& 8.56 & 7.33	& 7.79	& 8.01 & -- & -- \\
J1152-02	& 11:52:37	& -02:28:10	& -- & yes & -- &  -16.5	& 8.31 & 8.31	& 8.30	& 8.46 & 8.15 & -1.81 \\
J1225-06	& 12:25:40	& -06:33:07	& -- & yes & -- &  -14.0	& 8.48 & 7.61	& 7.75	& 8.00 & 7.50 & -1.82 \\
J1328+02	& 13:28:12	& +02:16:46	& -- & yes & -- &  -15.0	& 7.93 & 8.16	& 8.23	& 8.40 & 7.87 & -1.58 \\
J1403-27	& 14:03:35	& -27:16:47	& -- & yes & -- &  -16.3	& 8.72 & 8.16	& 8.20	& 8.38 & 7.94 & -1.70 \\
J1609-04	& 16:09:37	& -04:37:13	& -- & yes & -- &  -16.1	& 8.30 & 8.32	& 8.32	& 8.47 & 8.35 & -1.84 \\
J2039-63	& 20:38:57	& -63:46:16	& -- & yes & -- &  -16.0	& 8.31 & 8.35	& 8.32	& 8.49 & 7.97 & -1.67 \\
J2234-04	& 22:34:55	& -04:42:04	& -- & yes & -- &  -16.3	& 8.50 & 8.05	& 8.12	& 8.31 & 7.90 & -1.79 \\
J2242-06	& 22:42:24	& -06:50:10	& -- & yes & -- &  -15.5	& 7.95 & 8.03	& 8.11	& 8.30 & 7.92 & -1.81 \\
J2254-26	& 22:54:45	& -26:53:25	& -- & yes & -- &  -15.9	& 8.46 & 8.14	& 8.25	& 8.44 & 8.09 & -1.61 \\
J2311-42	& 23:11:11	& -42:50:51	& -- & yes & -- &  -16.3	& 8.19 & 8.19	& 8.24	& 8.42 & 8.09 & -1.61 \\
J2349-22	& 23:49:24	& -22:32:56	& -- & yes & -- &  -14.5	& 7.99 & 7.90	& 8.02	& 8.22 & 7.86 & -1.70 
\enddata
\end{deluxetable*}

MODS1 simultaneously observes a red and blue set up.  We used the G400L (400 lines mm$^{-1}$, R$\sim$1850) and G670L (250 lines mm$^{-1}$, R$\sim$2300) gratings to achieve broad spectral coverage from 3100 -- 10,000\,\AA. To minimize slit-loss from atmospheric refraction, we aligned the slit to the parrallactic angle of midpoint of each observation sequence, 3$\times$10 minute exposures per target.  Seeing was poor for Mt. Graham (1.2\arcsec -- 2\arcsec) with variable cirrus, thus we do not provide an absolute flux calibration.  Relative flux calibration between the red and blue setups was achieved by matching the continuum between [O\,\iii] $\lambda$5007 and H$\alpha$. The spectra were processed following Croxall et al. (in prep).  We fit the stellar continuum with models from \citet{bruzual2003} using STARLIGHT \citep{cidfernandes2005}.  When we could not constrain the underlying Balmer absorption with stellar models we adopt 2\AA\,EW absorption under H$\beta$.  We measure emission line fluxes by fitting them with Gaussians.  

We detect the strong [O~\iii], [O~\ii], H$\alpha$, and H$\beta$ lines (Table \ref{tab:lines}) in all galaxies.  In four galaxies we robustly detect the [N~\ii] $\lambda$6584\AA\ line while in three galaxies we also detect the faint auroral [O~\iii] $\lambda$4363\AA ~line.  We compute the gas-phase oxygen abundance using three strong-line calibrations, \citet[M91]{M91},  \citet[KK04]{KK04}, and \citet[PT05]{PT05}.  While the absolute calibration of these strong-line abundance indicators continues to be debated \citep[e.g.,][]{Kewley2008}, these three calibrations span the range of absolute scales and include both empirical calibrations and those based on photoionization models and the relative calibrations are reliable.  Where possible, we also calculate the electron temperature to determine an abundance using direct methods, though we note that these are not deep observations and the low-signal-to-noise limit will bias our observations towards systematically low metallicities.

Our dwarf galaxies all have an absolute magnitude, M$_r$ $>$ -16.2 (M$_B$ $>$ -16.0).   For comparison with the control samples, we calculate the B-band absolute magnitude by transforming the SDSS colors.  This is applicable to galaxies without strong emission lines, and may result in an overestimate in the B-band magnitude for some of our dwarfs.

We draw control samples from two studies: 21 isolated dwarf irregular galaxies from  \cite{vanZee2006}, and 13 isolated gas-rich dwarf galaxies from \cite{Nicholls2014b}.  While the isolation criteria varies slightly between the two samples, they generally result in galaxies with no neighbors within $\sim$200 kpc, suggesting they have not undergone a substantial interaction for $\sim$2 Gyr.  By similar criteria, we classify four void dwarf galaxies as isolated.  The remaining four void dwarf galaxies have close ($<$40 kpc) massive ($>$10$^8$ M$_\sun$) companions, all of which show signs of tidal interaction in the H~\textsc{i} maps (Figure \ref{fig:ha}).

As both control samples are within 30 Mpc, we cannot apply the same large-scale structure identification techniques as we have in the VGS \citep{Platen2007, Aragon-Calvo2010, Kreckel2011}, however the large-scale structure in the local universe is relatively well mapped \citep{Lavaux2010, Kitaura2012}.   Two galaxies (KK246 and HIPASS J1609-04) from the \cite{Nicholls2014b} sample fall within the local void (see also \citealt{Kreckel2011b, Nicholls2014a}), but the remaining galaxies from both samples do not fall within any of the established nearby voids \citep{Tikhonov2006, Tully2008, Elyiv2013}.  Due to the isolation criteria, we classify the remainder as `field' galaxies residing in average density walls or filaments.  

For all galaxies in the control sample, we use the measured line fluxes to recalculate the gas-phase metallicity in a manner consistent with our sample for the three strong-line methods.  While these strong-line methods can have large (0.7 dex) absolute uncertainties, relative uncertainties within each assumed calibration are $\sim$0.1 dex \citep{Kewley2008}, which dominates over measurement uncertainties for the strong lines.  

We do not expect aperture biases to significantly affect our comparison. We sample $\sim$500 pc scales, equivalent to the scales probed by \cite{Nicholls2014b} but larger than the $\sim$100 pc scales sampled by  \cite{vanZee2006}.  Most of the abundances calculated are based on measurements of only one or two HII regions in each galaxy.

\begin{deluxetable*}{c | c c c c c c c c c}[]
\tablecaption{Reddening corrected line fluxes relative to H$\beta$.
\label{tab:lines}}
\tablehead{
\colhead{Line} &
\colhead{VGS\_30a} &
\colhead{VGS\_34a} &
\colhead{VGS\_36a} &
\colhead{VGS\_38a} &
\colhead{VGS\_38b} &
\colhead{VGS\_39a} &
\colhead{VGS\_51a} &
\colhead{VGS\_56a}
}
\startdata
$[$O~\ii$]~\lambda$3727+3729 & 3.40$\pm$0.32& 2.67$\pm$0.29 & 1.93$\pm$0.02 & 1.49$\pm$0.02 & 1.62$\pm$0.01 & 0.97$\pm$0.02 & 1.40$\pm$0.27 & 4.33$\pm$0.18 \\
H$\gamma$                                & \ldots                 & \ldots                & 0.47$\pm$0.02 & 0.43$\pm$0.02 & 0.45$\pm$0.01 & 0.46$\pm$0.01 & \ldots                 & 0.34$\pm$0.10 \\
$[$O~\iii$]~\lambda$4363                & \ldots                & \ldots                & \ldots                 & 0.11$\pm$0.03 & 0.05$\pm$0.01 & 0.06$\pm$0.01 & \ldots                 & \ldots                 \\
$[$O~\iii$]~\lambda$4959                & \ldots                & 0.84$\pm$0.02 & 1.10$\pm$0.01 & 2.06$\pm$0.01 & 1.25$\pm$0.01 & 0.85$\pm$0.01 & 1.12$\pm$0.14 & 0.34$\pm$0.10 \\
$[$O~\iii$]~\lambda$5007                & 1.27$\pm$0.22 & 1.67$\pm$0.01 & 3.35$\pm$0.01 & 6.20$\pm$0.01 & 3.78$\pm$0.01 & 2.59$\pm$0.01 & 2.60$\pm$0.14 & 1.04$\pm$0.10 \\
H$\alpha$                                & 2.69$\pm$0.17 & 2.36$\pm$0.02 & 2.31$\pm$0.01 & 2.28$\pm$0.01 & 2.78$\pm$0.01 & 2.76$\pm$0.01 & 2.76$\pm$0.14 & 2.49$\pm$0.11 \\
$[$N~\ii$]~\lambda$6584          & \ldots                 & \ldots                 & 0.10$\pm$0.02 & 0.08$\pm$0.01 & 0.08$\pm$0.01 & 0.04$\pm$0.01 & \ldots                 &  \ldots                \\
EW(H$\beta$)                             & 6.59                  & 13.29  & 34.27                & 19.12               & 4.07                   & 117.3                & 37.3                  & 5.63
\enddata
\end{deluxetable*}

\section{Results}

\begin{figure}[b!]
\centering
\includegraphics[width=8.5cm]{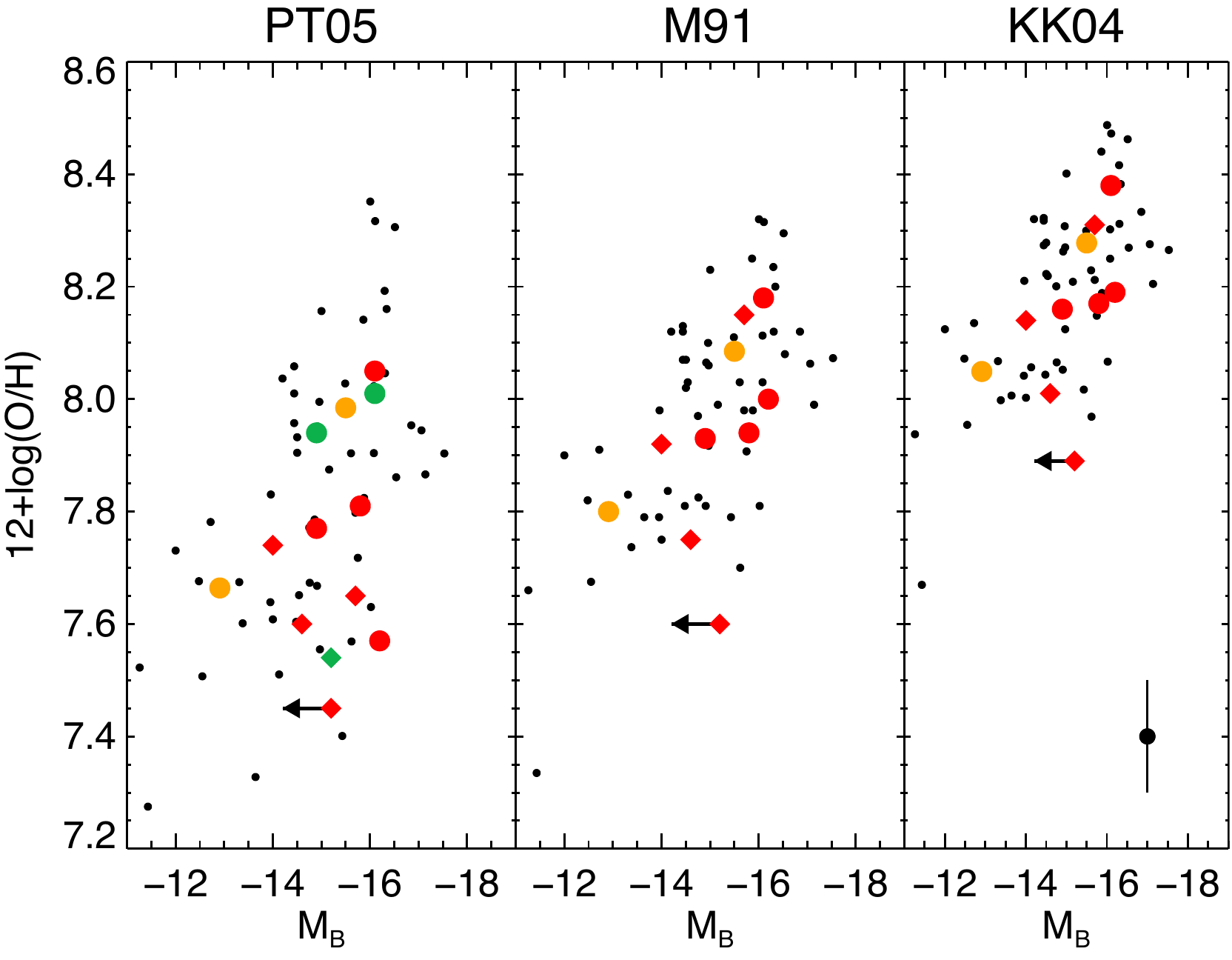}
\caption{
The gas-phase oxygen abundance as a function of the B-band absolute magnitude for VGS dwarfs (red) and isolated control sample galaxies (black) using the PT05, M91 and KK04 strong-line metallicity calibrations.  All show the established correlation for larger galaxies to have higher metallicity, though we span a narrow range in magnitude.  
Two of the \cite{Nicholls2014b} galaxies reside in voids (orange).  Three direct abundance measurements for the VGS dwarfs are  shown as compared to the PT05 results (green).  VGS dwarfs with nearby companions are marked with a diamond.  
We see no systematic offset in the metallicity of  void galaxies compared with the control, or between isolated and non-isolated dwarfs, in any of the calibrations.  VGS\_39a appears particularly metal deficient, however, we suspect that strong emission line contamination results in an overestimation of the B-band magnitude.   
\label{fig:zmb}}
\end{figure}

In all cases, our VGS dwarf galaxies have abundances consistent with the control sample, with no apparent systematic offset (Figure \ref{fig:zmb}).  This is consistent with the results of \cite{Nicholls2014a} for the two dwarfs within the Local Voids (also shown here).  

Four of the VGS dwarfs are not isolated, having a companion within $\sim$100 kpc in projection, most with further evidence for ongoing interaction in the extended H~\textsc{i} disks.  These four galaxies also do not show any offset relative to the isolated galaxy samples.  While this sample is too small to draw strong conclusions, it suggests that external gas accretion is playing a limited role in the chemical evolution of these systems, and that this evolution is instead dominated mainly by the internal secular processes that are linking the simultaneous growth and enrichment of these galaxies.

We note that one of the VGS dwarfs, VGS\_39a, exhibits a low metallicity for its estimated M$_B$, particularly in the M91 and KK04 diagnostics, and even with the temperature-sensitive method.  However, as this galaxy also exhibits a relatively large equivalent width in the [O~\iii] 5007\AA~ line (EW$_{5007}\sim220$\AA), we estimate that this line emission contributes 0.2 magnitudes to the measured g-band filter (which extends to 5500\AA) and invalidates our assumption of no strong emission lines in the conversion to B-band.   Using the SDSS photometry directly (see below) we estimate a stellar mass of 4 $\times$ 10$^7$ M$_\sun$, more consistent with M$_B \sim -14$ mag, making VGS\_39a less of an outlier in these plots. 

In addition to the established correlation between luminosity and metallicity, an additional dependence on star formation rate has also been shown \citep{Mannucci2010}.  This is not easy to explore in this sample, as the extrapolation from H$\alpha$ flux measured within individual HII regions to a global star formation rate is not straightforward.  However, previous \citep{Patiri2006, Park2007, Kreckel2012} and ongoing \citep{BeyguPhD} work studying the star formation within void galaxies shows there is no systematic difference compared to typical galaxies at fixed stellar mass, suggesting this additional parameter is unlikely to be affecting our result.

Another tracer of metal enrichment in galaxies is the ratio of nitrogen relative to oxygen.  
At high oxygen abundance, nitrogen is produced as a secondary element in intermediate mass stars, resulting in a correlation between N/O and O/H.  However, at low oxygen abundance (12+log(O/H) $<$ 8.5), the N/O ratio is approximately constant \citep{Lequeux1979}, suggesting nitrogen is produced mainly as a primary element in massive stars. Our void dwarf galaxies all fall within this low metallicity regime and exhibit N/O consistent with the fixed log(N/O) = $-$1.6 abundance ratio \citep{Thuan1995}, strengthening the conclusion that atypical chemical evolution is not observed in these void galaxies.  

\begin{figure}[h]
\centering
\includegraphics[width=9cm]{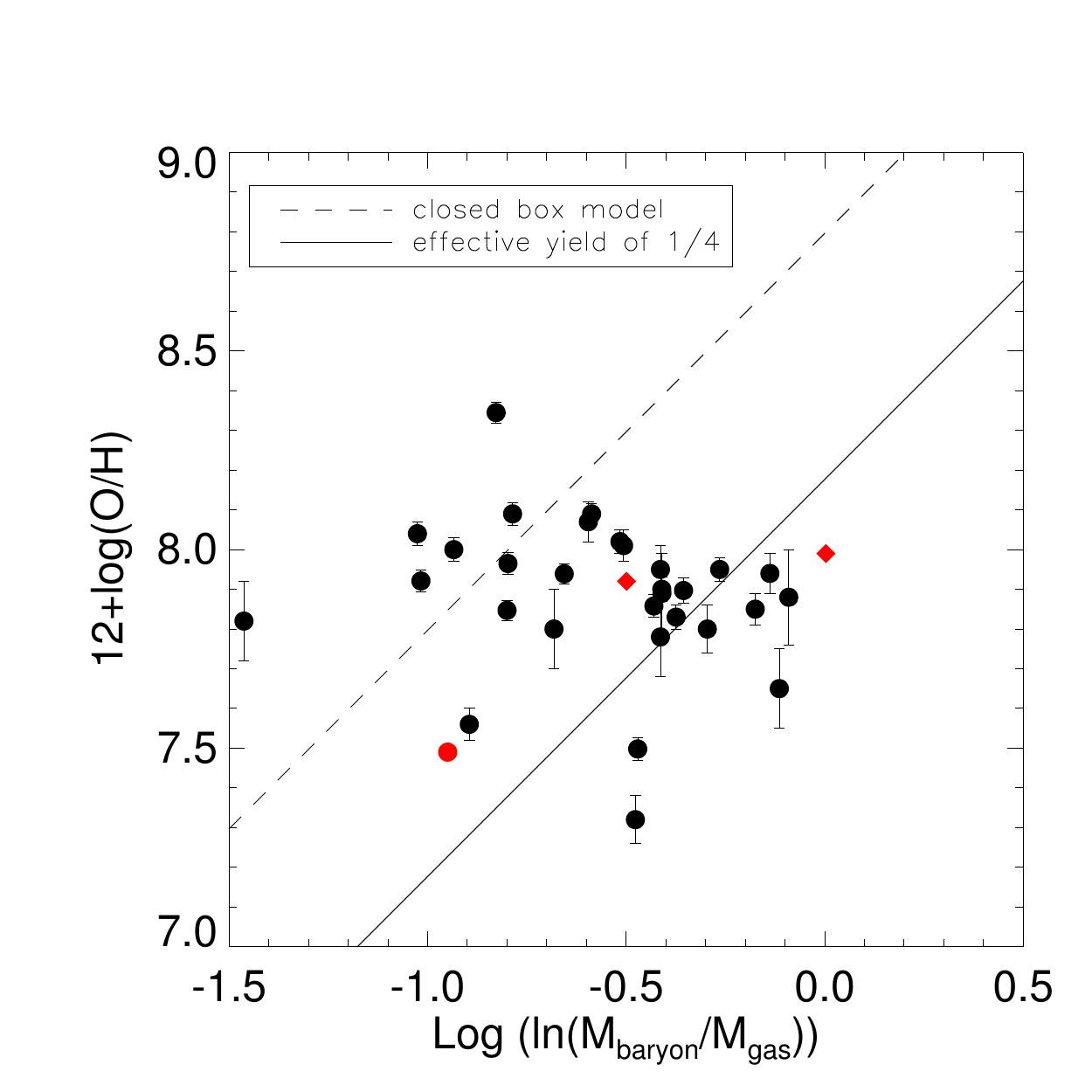}
\caption{Comparison of the metallicity calculated from direct temperature methods with the gas fraction, following \cite{vanZee2006}, for  VGS dwarfs (red) and control sample (black).   Two VGS dwarfs have nearby companions (diamonds).  
A closed-box evolutionary model prediction (dashed line) over-predicts the metallicity for most galaxies.  The void dwarf galaxies are in good agreement with the control sample, which shows a general effective yield 25\% of what is expected from a closed-box model.  
\label{fig:spec}}
\end{figure}

We examine the evolutionary state of our galaxies by comparing with a simple closed box model for the abundance evolution \citep{Edmunds1990}, which allows for no dilution by infalling pristine material or loss of metals by outflowing enriched material.  Following the assumptions outlined in \cite{vanZee2006}, we plot the metallicity as a function of gas fraction for a perfect closed box model (Figure \ref{fig:spec}).  
We estimate the stellar mass using the broadband photometry available from each sample and the conversions in \cite{Bell2003}.  For the \cite{Nicholls2014b} galaxies we use colors reported in \cite{Doyle2005}, and for the VGS dwarfs we use the SDSS photometry and colors directly.  The total gas mass is based on the H~\textsc{i} mass with a correction for neutral Helium (M$_{gas}$ = 1.3 M$_{HI}$).  We assume there is no significant molecular gas component in these systems, based on the low CO intensity typically found in dwarf galaxies \citep{Schruba2012}.  Here we have plotted only metallicities measured using direct methods, as taken directly from \cite{vanZee2006} and \cite{Nicholls2014b}, to reduce the scatter compared to the model. 

\cite{vanZee2006} found that the majority of dwarf galaxies are consistent with having an effective yield about 25\% that for a closed box system.  Our void dwarf galaxies are also consistent with this result.  Though we have only three void dwarf galaxies with direct abundances measured, we note that they exhibit both extremes in gas fraction,  suggesting significant scatter in the gas dynamics in voids.  We also note that two of these void dwarf galaxies are not isolated, and it is in fact one of these, VGS\_38a,  that falls furthest from the closed box model.  This galaxy is very clearly interacting with a larger nearby (D $\sim$ 40 kpc) companion (Figure \ref{fig:ha}).

\section{Discussion}

\subsection{Comparison with previous work}
We find no significant offset in gas-phase metallicity for our eight void dwarf galaxies with respect to the sample of isolated dwarf galaxies in average density environments (Figure \ref{fig:zmb}).  
\cite{Pustilnik2011}, however, found a more significant offset in their study of 48 void galaxies in the Lynx-Cancer void.  They found 12\% of their void galaxies demonstrate a large oxygen deficiency (12+log(O/H) $<$ 7.3), and the majority show a systematic offset to lower abundances by 0.15 dex.  We cannot precisely explain the difference between their result and ours, however there are a few key differences between the void galaxy samples that could play a role.

While their galaxy sample size is significantly larger, all galaxies are selected to reside within a single nearby void, and thus some may reside in sub-filaments within the void \citep{Sheth2004, Aragon-Calvo2013, Rieder2013, Alpaslan2014}.  Our sample is carefully selected to contain only galaxies residing centrally within different 10-20 Mpc diameter voids.  This does not rule out the possibility of filaments within our sample \citep{Beygu2013}, and in fact VGS\_38a and VGS\_38b both appear in a linear configuration with a third, larger void galaxy (VGS\_38).  These two dwarf galaxies, however, also appear indistinguishable from the other void dwarf galaxies and the isolated dwarf galaxy control sample.  

Many of the \cite{Pustilnik2011} sample are classified as low surface brightness (LSB) galaxies, which suggests a relatively unevolved state where star formation and interactions have not yet built up the galaxy disk.  
There is some theoretical expectation that collapse conditions  forming the most massive galaxies in voids will result in LSB extended disks \citep{Hoffman1992}, however this is not predicted to affect the dwarf galaxies as they do not arise from extreme initial perturbations in the density field.  SDSS suffers from a bias against selecting LSB galaxies in the redshift survey, making a careful study of the trends in voids difficult.  However, as our sample is purely H~\textsc{i} selected, it should not suffer from this bias.  
Only VGS\_30a could potentially be classified as LSB, though it is also clearly interacting with a neighbor  within 100~kpc (Figure \ref{fig:ha}).  The predominance of LSB galaxies in the \cite{Pustilnik2011} sample compared to ours could play a significant role in the offset they observe towards lower metallicity.

Finally, as our sample is H~\textsc{i} selected, we have a bias towards gas-rich galaxies that is not influencing the selection of Lynx-Cancer void galaxies \citep{Pustilnik2011b}.  As gas-rich dwarfs have the fuel more readily available to form stars and enrich the surrounding material, a gas-rich selected sample could be expected to exhibit higher metallicity.  A relation connecting stellar mass, metallicity and H~\textsc{i} mass has been found for more massive ($M_* > 10^9 M_\sun$) galaxies \citep{Bothwell2013}, though the trend is in the opposite direction - the more H~\textsc{i} rich galaxies at fixed stellar mass have lower metallicity.  

Our result is more consistent with the findings of \cite{Nicholls2014a}, who also used a H~\textsc{i} selected sample to examine two dwarf galaxies in the Local Void and found they have typical metallicities for their luminosity.  The abundance ratio of nitrogen to oxygen for HII regions in their two void galaxies are consistent with the lower values we find.  

\subsection{Implications for dwarf galaxy evolution}
The discrepancy between observed dwarf galaxies and the evolution prescribed by a close-box model requires either the inflow of pristine material or outflow of enriched material.  These mechanisms are also thought to regulate the development of the metallicity to luminosity (or stellar mass) relation observed over the full range in galaxy stellar mass \citep{Skillman1989, Tremonti2004}. 

Detailed study of the H~\textsc{i} morphology and kinematics in void galaxies shows evidence of ongoing accretion of gas  \citep{Kreckel2011b, Kreckel2012}, potentially through a cold mode of accretion \citep{Stanonik2009}.  Cold accretion, in particular, should enable pristine gas to penetrate directly to the central star-forming region of the galaxy disk, and be immediately available to form stars.  This accretion mode is seen in simulations to play an important role in low mass halos, which are preferentially found both at high redshift and in low density environments at z=0 \citep{Keres2005}.  

If we believe cold accretion is more prevalent in voids, then the lack of offset within the metallicity-luminosity relation in our void dwarf galaxy sample suggests inflows alone do not play a significant role in driving the metallicity evolution of dwarf galaxies.   
Some evidence for this may be seen directly in KK246, a dwarf galaxy in the Local Void, which shows evidence of gas accretion \citep{Kreckel2011b} yet has a metallicity typical for its size \citep{Nicholls2014a}.  
However, the assumption that gas accretion is directly available to form stars may fail in the case where accreted gas forms an extended disk, well beyond the optical disk where star formation is observed to occur.  In KK246, the H~\textsc{i} extends out to five times the optical radius.  As the H~\textsc{i} detected in the eight galaxies presented in this letter is largely unresolved (Figure \ref{fig:ha}) we cannot comment directly on the role the distribution of the H~\textsc{i} plays in these galaxies.

Additionally, four of the eight void dwarf galaxies have massive nearby companions that are clearly interacting in H~\textsc{i}.  It may be expected that these companions pre-enrich the gas in-falling onto the neighboring dwarfs \citep{Croxall2009}, however, again we see no systematic difference between these and the isolated void dwarf galaxies.  In fact VGS\_30a, which is only 17~kpc from a massive companion, shows the smallest effective yield relative to the closed box model (Figure \ref{fig:spec}).  

These results suggest it is more likely the outflow of enriched material that drives the metallicity-luminosity relation.  As this is an internal process  operating independently of the surrounding large-scale environment, it is reasonable that the majority of void galaxies will not show a significant deviation from the established relation.  This may not hold true in specific cases, such as in low surface brightness void galaxies \citep{Pustilnik2011} where the star formation may be more distributed and thus less efficient in driving outflows.  This may also fail in cases where the surrounding environment is more dense, such as clusters or groups.

We find that the gas-phase metallicity of these void dwarf galaxies does not differ significantly from typical dwarf galaxies at fixed luminosity, this alone is a surprising result, as it indicates that galaxy evolution proceeds largely independent of significant differences in the surrounding large-scale environment.  In this way, void galaxies provide a well constrained population of galaxies that are ideal for detailed study of the internal physical processes that drive galaxy evolution.

\acknowledgements
We thank Thijs van der Hulst for valuable comments.  KK acknowledges grants GR 3948/1-1 and SCHI 536/8-1 from DFG Priority Program 1573.  
KC is grateful for support from NSF Grant AST-6009233, and visitor funding from DFG Priority Program 1573.  
This paper uses data taken with the MODS spectrographs built with funding from NSF grant AST-9987045 and the NSF Telescope System Instrumentation Program (TSIP), with additional funds from the Ohio Board of Regents and the Ohio State University Office of Research.  
This paper made use of the modsIDL spectral data reduction pipeline developed in part with funds provided by NSF Grant AST-1108693.  
This work was based in part on observations made with the Large Binocular Telescope (LBT). 
Funding for SDSS-III has been provided by the Alfred P. Sloan Foundation, the Participating Institutions, the National Science Foundation, and the U.S. Department of Energy Office of Science. 



\end{document}